\documentclass[10pt,aps,prl,twocolumn,superscriptaddress,floatfix,longbibliography]{revtex4-1}

\usepackage{graphicx}
\usepackage{amsfonts}
\usepackage{amssymb}
\usepackage{amsmath}
\usepackage{txfonts}
\usepackage{lipsum}
\usepackage[dvipsnames]{xcolor}
\usepackage{wasysym}
\usepackage[colorlinks=true, allcolors=blue]{hyperref}
\usepackage{bbold}
\usepackage{etoolbox} 
\usepackage[normalem]{ulem}
\usepackage{mathtools} 
\usepackage{multirow} 
\usepackage{hhline}
\usepackage{mathrsfs} 
\usepackage{soul}
\usepackage{array}

\usepackage{letltxmacro}
\LetLtxMacro{\oldsqrt}{\sqrt}
\renewcommand{\sqrt}[2][\mkern8mu]{\mkern-6mu\mathop{}\oldsqrt[#1]{#2}}

\DeclareMathOperator{\Tr}{Tr}

\begin{document}

\title{Superconductivity of Bad Fermions: Origin of Two Gaps in HTSC Cuprates}

\author{Evgeny A. Stepanov}
\affiliation{CPHT, CNRS, {\'E}cole polytechnique, Institut Polytechnique de Paris, 91120 Palaiseau, France}
\affiliation{Coll{\`e}ge de France, Universit{\'e} PSL, 11 place Marcelin Berthelot, 75005 Paris, France}

\author{Sergei Iskakov}
\affiliation{Department of Physics, University of Michigan, Ann Arbor, Michigan 48109, USA}

\author{Mikhail I. Katsnelson}
\affiliation{Institute for Molecules and Materials, Radboud University, 6525AJ Nijmegen, The Netherlands}
\affiliation{Constructor Knowledge Institute, Constructor University, 28759 Bremen, Germany}

\author{Alexander I. Lichtenstein}
\email{alexander.lichtenstein@uni-hamburg.de}
\affiliation{Institute of Theoretical Physics, University of Hamburg, 20355 Hamburg, Germany}
\affiliation{European X-Ray Free-Electron Laser Facility, Holzkoppel 4, 22869 Schenefeld, Germany}

\begin{abstract}
We investigate the spectral properties of the doped ${t\text{\,-\,}t'}$ Hubbard model with parameters typical for high-temperature cuprate superconductors. 
Our approach is based on a novel strong-coupling Green's function expansion around a reference system -- the exactly solvable undoped particle-hole symmetric Hubbard lattice -- that possesses a large antiferromagnetic Mott-Hubbard-Slater gap in the electron spectrum.
The electron spectral function in the case of a large next-nearest-neighbor hopping ${t'=-0.3t}$, which is characteristic of the ${T_c \approx 100\,\text{K}}$ family of cuprates, reveals a strongly renormalized flat band feature with a pseudogap around the antinodal point.
The superconducting response of this system to a small ${d_{x^2-y^2}}$-like external field exhibits a very unusual form. 
It features a pseudogap at the antinodal point in the normal part of the Nambu Green's function, related to a ``bad-fermion'' behavior in a normal phase, as well as a ${d}$-wave-like structure in the anomalous (Gorkov's) Green's function, with zero response at the nodal point of the Brillouin zone. 
Remarkably, we find that the anomalous part of the response deviates essentially from the simplest ${(\cos{k_x}-\cos{k_y})}$ form in momentum space. 
Specifically, its extrema are shifted away from the ${(\pi,0)}$ and ${(0,\pi)}$ points due to suppression of the response by the pseudogap.
The observed two-gap structure of the electron spectra in a generic strong-coupling model of cuprates can serve as a basis for phenomenological treatment of different physical properties of high-temperature superconductors within two-fluid model.
\end{abstract}

\maketitle

The high-temperature superconductivity (HTSC) in copper-oxide-based perosvkites was discovered almost 40 years ago~\cite{Muller_HTSC}, but a clear and complete theoretical view on this remarkable effect is still missing. 
Regarding the generic electronic structure of HTSC cuprates, there is a consensus on the paramagnetic Fermi surface: the angle resolved photoemission spectrum (ARPES)~\cite{ARPES_ZX} agrees well with the results of the Density Functional Theory (DFT)~\cite{OKA_TB}. 
The main feature, related to a single Cu${3d}$-O${2p}$ band per plane crossing the Fermi energy, is described by a simple tight-binding model with a large next-nearest-neighbor (NNN) hopping $t'$. 
The typical ratio of $t'$ to the nearest-neighbor (NN) hopping amplitude $t$ is ${t'/t \simeq -0.3}$ for bismuth strontium calcium copper oxide ``BSCCO'' or yttrium barium copper oxide ``YBCO'' with a critical temperature on the order of ${T_c \approx 100\,\text{K}}$~\cite{ARPES_ZX}.
The sign and large values of $t'$ originate from the DFT band structure, which accurately describes the general chemical bonds in YBCO-like systems and is related to the strong hybridization of the valence band with the Cu$-{4s}$ orbital~\cite{OKA_TB}. 
Furthermore, the estimation of ${t'/t}$ for different HTSC cuprates~\cite{Pavarini} shows strong empirical correlations between this parameter and the $T_c$ values. 
Additionally, the negative sign of ${t'/t}$ leads to a shift of the van Hove singularity towards the chemical potential for optimal hole doping, which efficiently increases the critical transition temperature. According to DFT calculations~\cite{Novikov96}, the Fermi energy goes closer to van Hove singularity simultaneously to the growth of $T_c$ in mercury barium copper superconductor under pressure, the system with the highest $T_c$ among cuprates.  

\begin{figure}[t!]
\centering
\includegraphics[width=0.85\linewidth]{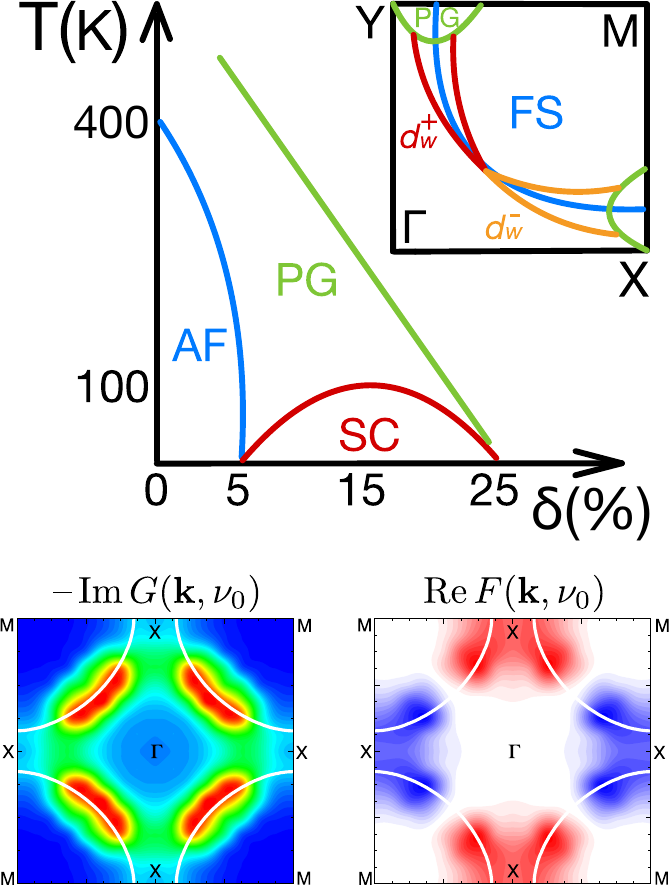}
\caption{Top panel: Generic phase diagram of HTSC cuprates as a function of temperature $T$ and hole doping $\delta$, and the corresponding Fermi surface (FS) with two distinct gap structures.
Bottom panels: Normal $G({\bf k})$ (left) and anomalous $F({\bf k})$ (right) parts of the Nambu-Gor'kov Green's function obtained at the zeroth Matsubara frequency ${\nu_0=\pi{}T}$ using the DF lattice DQMC scheme. The Green's functions are calculated for ${\delta=12\%}$ hole doping of the $t$\,-\,$t'$ Hubbard model. The non-interacting Fermi-surface is depicted by a white line.}
\label{fig:HTSCview}
\end{figure}

The energy spectrum observed in ARPES compared to the DFT bands reveals a strong many-body renormalization~\cite{ARPES_ZX}. 
The most unusual features of ARPES experiments~\cite{Fujimori_ARPES, ZX_2Gaps, Kondo_2Gap_2009, Hashimoto_2Gaps, Kordyuk_3G, Campuzano2G, Hashimoto2014, He_ZX_2G}, are related to the structure of the superconducting gap. 
The key findings of these experiments, specifically the emergence of two distinct gaps, are schematically depicted in Fig.~\ref{fig:HTSCview}, alongside the generic phase diagram of cuprates in the temperature $T$ {\it vs.} hole doping $\delta$ plane.
A growing body of experimental evidences suggests that the pseudogap phase terminates at the same doping as the superconducting dome~\cite{Davis2008, Taillefer_rev, He_ZX_2G, ZX_2Gaps, Campuzano2G}. 
It is therefore natural to propose that the driving mechanism for $d$-wave superconductivity in cuprates is connected to the formation of ``bad fermions'' near the antinodal regions in the normal phase, which are responsible for the pseudogap. 
Furthermore, this region of the Fermi surface is consistently associated with the pseudogap (Fig.~\ref{fig:HTSCview}), as evidenced by spin susceptibility measurements~\cite{Keimer2024} and scanning tunneling microscopy (STM) results~\cite{Davis2008,Boyer_2Gaps_STM}. Note also appearance of the second line in~\cite{Campuzano2G}, coherence temperature growing with the hole concentration which means that we probably deal with the coexistence of ``bad'' and ``good'' (coherent) fermions, as was also discussed based on transport properties~\cite{Ayres22}. Our theoretical consideration seems to confirm the experimental picture~\cite{Campuzano2G,Ayres22}; note that at the qualitative level a similar view was discussed in the last work of P. W. Anderson~\cite{swan}.     

From a theoretical perspective, it has been shown using numerically exact density matrix renormalization group (DMRG) methods and the constrained-path determinant Quantum Monte Carlo (DQMC) scheme that the single-band Hubbard model with ${t' = 0}$ does not exhibit superconductivity in its ground state for any parameter values~\cite{Zhang2020}. 
Instead, it displays a stripe phase at low temperatures.
However, introducing NNN hopping ${t'}$ leads to the emergence of $d$-wave superconductivity along with a partially filled stripe order for moderate values of ${t' = -0.2t}$~\cite{Zhang2024}, which correspond to LSCO-like systems~\cite{Pavarini}.
For larger values of ${t' = -0.25t}$, $d$-wave superconductivity becomes the dominant instability in the single-band Hubbard model~\cite{Devereaux2019}. 

In turn, variational QMC calculations provide conflicting results regarding the effect of $t'$ on superconductivity in HTSC cuprates.
While recent studies of realistic one-band model of different compounds~\cite{Imada2023} find no correlations with $t'$ and a monotonic increase of pairing with $U$, a similar calculations for the simple $t$\,-\,$t'$ model~\cite{Ogata2013} find strong evidence for a two-gap structure with strong pairing, for ${t'\leq 0}$ and an optimal value of ${U\simeq 7t}$.
The latter calculations also reveal ``bad fermion'' behavior near the antinodal point for a certain doping range, meaning that the occupation number $n(\mathbf{k})$ deviates significantly from the Fermi step function, and this effect turns out to be very sensitive to ${t'/t}$.
However, this approach cannot access frequency-dependent properties, providing limited insight into electron dynamics.

The first successful application of cluster extension of the Dynamical Mean Field Theory (C-DMFT) to study the interplay between antiferromagnetism (AFM) and superconductivity in cuprates~\cite{dwCDMFT}, along with numerous subsequent efforts in the same direction, faced challenges related to a poor momentum-space resolution and artificial breaking of translational symmetry unavoidable after separation of a cluster from the full crystal lattice. However, cluster approaches remain valuable for semi-quantitative analysis of peculiarities in the electronic structure by connecting them to the quantum states of individual plaquettes. This connection, in particular, explains high sensitivity of the results to the ${t'/t}$ ratio and uncovers a useful relations to Kondo lattice systems due to nontrivial degeneracy in plaquettes~\cite{Harland16, Harland20, Danilov2022}.

An important finding from C-DMFT calculations~\cite{Millis50} is that more than ${50\%}$ of superconducting pairing in the single-band Hubbard model with optimal values of parameters arises from strong-coupling effects beyond the standard spin-fluctuation mechanism within BCS-like theories~\cite{Scalapino12}. At the same time,
it is crucial to emphasize that the large interaction limit, where the Hubbard interaction ${U \gg W}$ (with the bandwidth ${W = 8t}$), is equivalent to the $t$\,-\,$J$ model and fails to accurately describe cuprate physics. 
In this limit, $d$-wave superconductivity in the $t$\,-\,$J$ model only emerges in the ``unphysical'' regime of ${t'/t > 0}$~\cite{White2021, Sheng2021, Gong2024}. 
Furthermore, the large-$U$ limit is unfavorable for two-hole pairing energy, as demonstrated by exact diagonalization of a small ${4 \times 4}$ periodic system~\cite{Danilov2022}.

We would like to point out an important synergy effect with standard spin-fluctuation mechanism of $d$-wave superconductivity~\cite{Millis90, Scalapino12} and effects of van-Hove singularity which provides band flattening and nontrivial interplay of instabilities in particle-particle and particle-hole channels~\cite{IKK2001, IKK2002}. 
Different investigations
shows that for optimal doping, Coulomb correlation parameter ${U/t\simeq5.6}$ and ${t'/t\simeq -0.3}$ the system is very close to Lifshitz 
transition with coincidence of van-Hove singularity with chemical potential~\cite{Ferrero2018, Harland20} which drastically boosts
spin-fluctuations and together with strong-coupling effects favors high-temperature superconductivity.
An important connection between topologically ordered Higgs phase related to SU(2) local moment theory and pseudogap phenomena in cuprate put forward~\cite{Sachdev2018} and confirmed by eight-site momentum space C-DMFT calculations~\cite{Ferrero2018}. More elaborated investigation of pseudogap states using recently developed diagammatic Monte-Carlo method~\cite{Georges2024} also proposes a connection to the ground-state stripe phase calculated at zero temperature. 

What is still missing in the HTSC puzzle is the understanding of nontrivial relation of pseudogap phenomena to \mbox{$d$-wave} superconductivity.
From the one hand, both effects are related to antiferromagnetic Mott phase of a parent undoped compound.
From the other hand, there is a competition between them for antinodal region. All this makes theoretical description of 
a generic HTSC phase diagram (Fig.~\ref{fig:HTSCview}) largely unsolved problem.

In this paper, we employ a novel Green's function strong-coupling lattice perturbation theory~\cite{DFQMC} to investigate the spectral properties of the doped single-band \mbox{$t$\,-\,$t'$} Hubbard model in the parameter regime relevant for HTSC cuprates.
We find that at optimal doping, the electronic spectral function exhibits a pseudogap at the antinodal point in the Brillouin zone (BZ).
This pseudogap coexists with a superconducting response to an external \mbox{$d$-wave} field, which displays $d_{x^2-y^2}$ symmetry in momentum space.
Since the pseudogap is present already at temperature higher than the magnetic ordering one, extrapolated for ${\delta \to 0}$, we propose that the emergence of this ``bad fermion'' behavior at the antinodal point is directly linked to the appearance of local magnetic moments (LMMs) and their AFM correlations, constituting the key microscopic mechanism of high-temperature superconductivity.
The LMMs are mainly formed by the quasi-localized heavy ``bad fermions'', which is a dynamical process, while the region near the nodal point remains metallic, with well-defined quasiparticle states near the Fermi surface.
This dichotomy between the nodal and antonidal points in the spectral function is a characteristic feature of the intermediate regime of electronic correlations, where the strong-coupling physics associated with LMMs coexists with itinerant electronic behavior~\cite{PhysRevLett.132.236504}.
We further support this claim by explicitly disentangling the contributions of spin fluctuations of purely itinerant order (paramagnons) and the LMMs, introduced via an effective Higgs field, to the electronic self-energy within a single-site-based strong-coupling approach.
Our analysis reveals that while the presence of the pseudogap in the electronic spectrum suppresses superconductivity, the AFM correlations between the LMMs that give rise to this pseudogap simultaneously play a crucial role in enhancing the superconducting response.

\section*{Results}
\vspace{-7pt}
To investigate the two-dimensional one-band model for the Cu-O plane, we consider the $t$\,-\,$t'$ Hubbard model on a ${16 \times 16 \times 64}$ space-time grid (periodic in space, antiperiodic in imaginary time) with the NN ${t = 1}$ (which defines the unit of energy) and NNN ${t' = -0.3}$ hopping amplitudes, and a local (on-site) Coulomb interaction ${U = 5.6}$. 
The choice of the interaction strength is based on a highly degenerate point in energy spectrum observed in small cluster calculations, which favors superconductivity~\cite{Harland16, Harland20, Danilov2022}. 
The same value of $U$, related to the optimal nodal-antinodal dichotomy near the Lifshitz transition, was reported in Ref.~\cite{Wei_point}. 
Results for other lattice sizes further support the main conclusions of this work and are available in the Supplemental Material.

To solve the considered model, we employ a novel strong-coupling perturbative scheme that performs a diagrammatic expansion based on the interacting lattice problem~\cite{DFQMC}. 
Importantly, the reference problem is introduced on the same ${16 \times 16 \times 64}$ space-time grid as the original model and accounts for the same value of the Coulomb interaction $U$. 
However, the reference problem is considered particle-hole symmetric, i.e., at half-filling with ${t' = 0}$, which enables its efficient numerically exact solution using the DQMC scheme, as it turns out to be free of fermionic sign problem thanks to the particle-hole symmetry.
The diagrammatic expansion beyond the reference problem is carried out in the spirit of the Dual Fermion (DF) approach~\cite{RKL08, BRENER2020168310}. 
The next-nearest-neighbor hopping $t'$ and the difference between the chemical potentials of the reference and original systems $\mu$ serve as perturbative parameters for this expansion. 
This approach allows us to start from a correlated antiferromagnetic Mott insulator as the reference state and investigate doped pseudo-gap metals with strong $d$-wave-like superconducting fluctuations.
Since the difference in chemical potential is of the order of ${\mu = -1.5}$ for 15\% hole doping, and the maximum contribution of the NNN hopping ${4t' = -1.2}$ is small compared to the bandwidth ${W = 8t}$, while the simulation temperature is not very low, we consider the first-order DF diagrammatic correction to be sufficient for reasonably large lattice systems (see Materials and Methods).
Note that the diagrammatic expansion is formulated in momentum ${\bf k}$, Matsubara frequency $\nu$~\cite{AGD}, and ${2\times2}$ Nambu space with normal (diagonal) $G({\bf k},\nu)$ and anomalous (off-diagonal) $F({\bf k},\nu)$ contributions to the Nambu-Gor'kov Green's function. 

\begin{figure}[t!]
\centering
\includegraphics[width=1.0\linewidth]{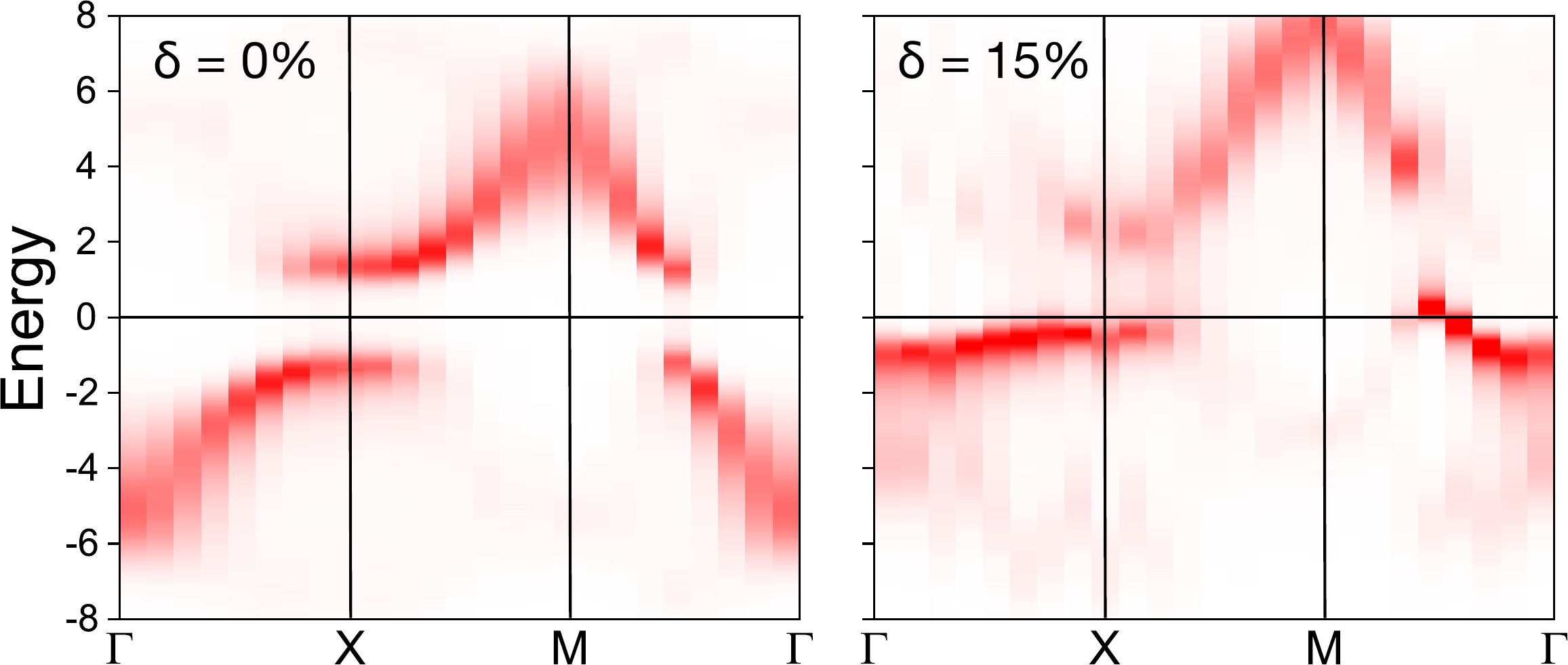}
\caption{The electronic spectral function of the $t$\,-\,$t'$ Hubbard model calculated for ${U=5.6}$ at ${\beta=5}$. The results are obtained for the half-filled particle-hole symmetric case (${t'=0}$, left panel) and for ${\delta=15\%}$ hole doping with ${t'=-0.3}$ (right panel). }
\label{fig:Ak}
\end{figure}

In Fig.~\ref{fig:Ak} we show the transformation of the electronic spectral function from the half-filled (reference) case with ${\mu_0=0}$ and ${t'=0}$ (left panel) to the doped system with ${\mu=-1.45}$ and ${t'=-0.3}$ (right panel).
The results are calculated at the inverse temperature ${\beta=1/T=5}$ using stochastic analytical continuation from Matsubara space to real energy~\cite{SOM2}. 
In the Mott insulator phase, corresponding to ${\delta=0\%}$ doping, one can see the formation of broad Hubbard bands around the energy ${E=\pm6}$, and shadow antiferromagnetic bands at ${E\simeq-4}$ in the vicinity of the $M=(\pi,\pi)$ point. 
Upon ${\delta=15\%}$ hole doping, the spectral function changes dramatically.
One can clearly see a strong effect of $t'$ on the van Hove singularity that results in the formation of a narrow, almost flat band in the $\Gamma$-X direction and the appearance of a pseudogap near the X point, which signals the quasi-localized behavior of electrons related to the formation of LMMs.
On the other hand, the spectrum remains metallic near the nodal point $(\Gamma-\text{M})/2$. 
Flattening of the bands and enhancement of van Hove singularities near the Fermi energy due to correlation effects were earlier found and studied by weak-coupling renormalization group~\cite{IKK2002} and strong-coupling DF approach~\cite{Yudin2014}. 

\begin{figure}[t!]
\centering
\includegraphics[width=0.7\linewidth]{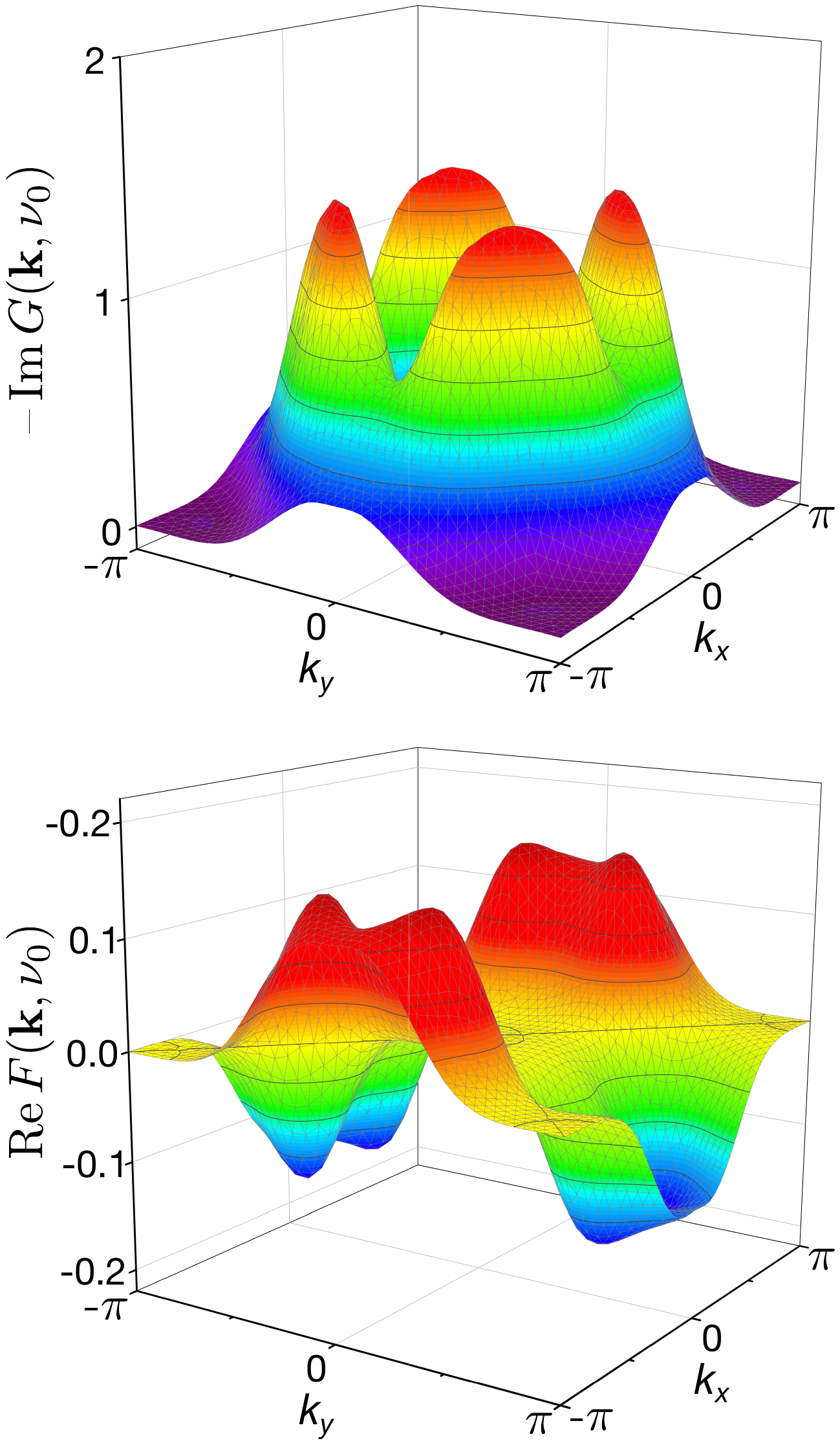}
\caption{The imaginary part of the normal Green function $G$ (top panel) and the real part of the anomalous Green function $F$ (bottom panel) calculated at the lowest Matsubara frequency ${\nu_0=\pi/\beta}$ for 12\% hole doping and ${\beta=8}$.}
\label{fig:Gk}
\end{figure}

In Fig.~\ref{fig:Gk} we plot the imaginary part of the normal Green's function $G({\bf k})$ (top panel), as well as the real part of the anomalous Green's function $F({\bf k})$ (bottom panel).
The results are obtained at the lowest Matsubara frequency $\nu_0=\pi{}T$ for the first BZ in the presence of a small external superconducting $d$-wave field with the amplitude ${h_{dw}=0.05}$. 
The 2D projection of these Green's functions on the BZ plane is shown in bottom panels of Fig.~\ref{fig:HTSCview}.
These calculations clearly capture the formation of a large pseudogap in the electronic spectral function ${-\frac{1}{\pi}\text{Im}\,G({\bf k})}$ at the antinodal $X=(\pi,0)$ point, which exists at already relatively high temperatures (${\beta=5}$ corresponds to ${T\simeq700\,\text{K}}$ for the realistic hopping amplitude ${t=0.3}$\,eV).
Additionally, we find that the anomalous Green's function $F({\bf k})$ is relatively large and has a very unusual shape.
Indeed, ${\text{Re}F({\bf k})}$ features a suppressed spectral weight at the X points, related to the pseudogap formation in the spectral function, which shifts its extrema in the direction toward the nodal point. 
We attribute such a strong deviation of the anomalous Green's function from a usual $(\cos{k_x}-\cos{k_y})$ form of an applied external $d_{x^2-y^2}$ field to a fingerprint of a strongly-correlated superconductivity.

\section*{Discussion}
\vspace{-7pt}
The obtained results call for identifying the microscopic processes behind the formation of the two-gap structure in the superconducting spectral function and understanding the implications of these processes for the mechanism of the strong-coupling theory of $d$-wave superconductivity.
Our general answer to this question is that this mechanism is similar to Anderson RVB~\cite{Anderson2004} and kinetic $t_ \bot$ mechanism of multilayer pair-hole hopping~\cite{Anderson1993, Anderson1998}. 
The main idea is that, informally speaking, it is not energetically favorable for fermions in normal phase to be ``bad'', that is, non-quasiparticle, so that they prefer to form, instead, a superconducting condensate. 
Anderson connected this energy balance with the interlayer hopping which is more difficult for non-Fermi-liquid state than for regular Fermi liquid. 
This seems to be not supported by further experimental development, and recent discovery of monolayer high-temperature cuprate superconductors~\cite{HTSC_1layer} probably closes the discussion. 
Contrary, we believe that this is in-plane NNN kinetic energy, which plays the role that Anderson attributed to the interlayer hopping. Despite we show that the pseudogap formation by itself competes with superconductivity, this is not the main effect of the NNN hopping. 
As was have demonstrated by exact diagonalization for a small, ${4 \times 4}$ periodic system, ${t'=-0.3}$ leads to a huge enhancement of the binding energy fo two holes, and this is related to the peculiarities of energy spectrum of isolated plaquette. We believe that this essentially strong-coupling physics provides the necessary missing component of superconducting glue. 
Indeed, in our point of view, the preformed superconducting pairs already exist in the small clusters of the order of ${4 \times 4}$ due to strong AFM correlations, and it is $t'$ that makes these hole pairs ``coherent'' even in the single Cu-O plane.
Another important effect comes from the large density of state at high-temperatures near the Fermi level for hole doping around ${\delta \simeq 15 \%}$ and ${t'=-0.3}$, which is related to a ``highly degenerate'' ground state of a small cluster~\cite{Danilov2022}. 
In this case, the formation of a pseudogap we can attribute to the physics of periodic Kondo problem or ``destructive interference phenomena''~\cite{Gunnarsson_2014}.
It would be very interesting to connect our microscopic approach to the phenomenological two-fluid model of cuprate superconductors~\cite{Ayres22}, which assumes the coexistence of two fermionic subsystems in the normal phase, a Fermi-liquid and a non-Fermi-liquid ones, where only the latter becomes superconducting at low temperatures.
For the large doping $\delta \geq 25 \%$ our DF lattice DQMC scheme shows that the "bad electrons" become much more coherent 
correlated quasiparticles (see Supplemental Material).

To identify the precise mechanism by which ``bad'' electrons become superconducting, we perform additional calculations using the triply irreducible local expansion (\mbox{D-TRILEX}) method~\cite{PhysRevB.100.205115, PhysRevB.103.245123, 10.21468/SciPostPhys.13.2.036}.
This method is similar to the previously introduced DF lattice DQMC scheme, as it also employs a strong-coupling diagrammatic expansion based on an interacting reference problem.
Compared to the DF lattice DQMC scheme, \mbox{D-TRILEX} accounts for more diagrammatic contributions to the self-energy by including an infinite set of ladder-type diagrams, enabling an accurate treatment of collective charge and paramagnetic spin fluctuations (paramagnons) with arbitrary spatial range.
The trade-off, however, is that this approach cannot handle large lattice reference problems as efficiently as the DF lattice DQMC scheme.
In this work, we use a single-site DMFT impurity problem as the reference system for \mbox{D-TRILEX} calculations, which are performed for ${64\times64}$ ${\bf k}$-points in the Brillouin zone and 128 Matsubara frequencies.

\begin{figure}[t!]
\centering
\includegraphics[width=1.0\linewidth]{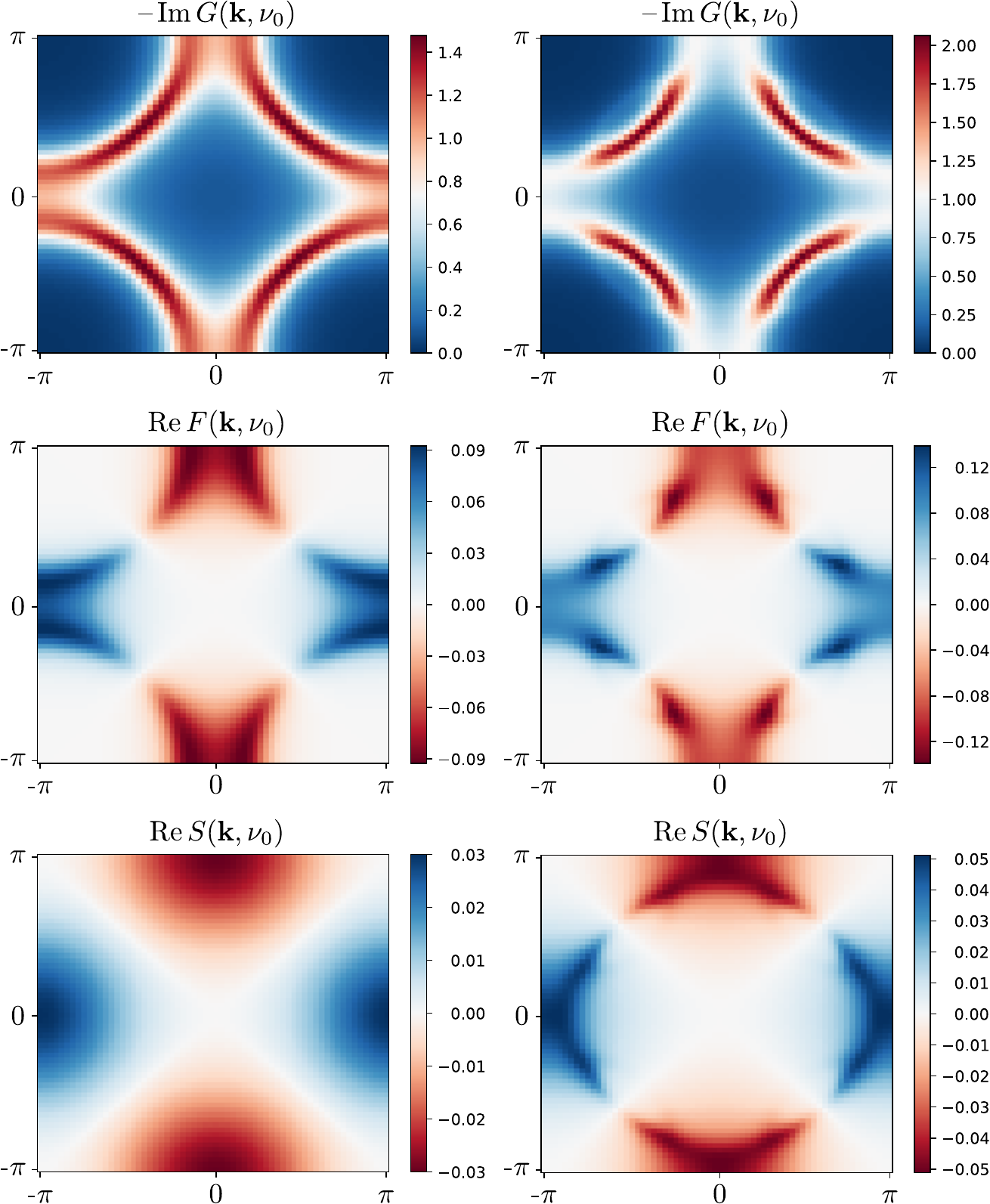}
\caption{Top panels: The imaginary part of the normal Green's function $G$. Middle panels: The real part of the anomalous Green's function $F$. Bottom panels: The real part of the anomalous self-energy $S$. The results are obtained using \mbox{D-TRILEX} at ${\beta=10}$ for ${\delta=16\%}$ of hall doping at the lowest Matsubara frequency ${\nu_0=\pi/\beta}$ as a function of momentum ${\bf k}$, comparing cases without (left column) and with (right column) the effective Higgs field ${h_{\rm AFM}=0.485}$.
}
\label{fig:PG}
\end{figure}

In this work, the \mbox{D-TRILEX} method serves as a complementary tool to provide deeper insights into the obtained results, as it enables the separation of different contributions to the self-energy.
In Figure~\ref{fig:PG}, we present the results obtained using \mbox{D-TRILEX} for the same model parameters as those in the DF lattice DQMC scheme, obtained at ${\beta=10}$ for ${\delta=16\%}$ hole doping and the ${h_{dw}=0.01}$ value of the superconducting $d$-wave field.
The left column shows the momentum-resolved normal $G({\bf k})$ (top panel) and anomalous $F({\bf k})$ (middle panel) contributions to the Nambu-Gor'kov lattice Green’s function, as well as the anomalous lattice self-energy $S({\bf k})$ (bottom panel), all obtained for the lowest Matsubara frequency.
At this relatively large temperature, we find that the spatial magnetic fluctuations are still relatively weak and do not lead to the development of a pseudogap in the normal Green’s function $G$.
The form of the anomalous Green’s function $F$ in momentum space resembles the normal Green’s function $G$, but with an additional $d$-wave-like symmetry, displaying zero response at the nodal point. 
The anomalous self-energy also exhibits a $d$-wave-like symmetry in momentum space, but in contrast to the Green’s function $F$, it reveals a ${(\cos k_x - \cos k_y)}$ profile that closely resembles the form of the applied external $d$-wave superconducting field (see Supplemental Material for details).

The results of the DF lattice DQMC scheme (Fig.~\ref{fig:HTSCview}), which clearly exhibit the pseudogap feature in $G$ and a substantial reduction in the intensity of $F$ at the antinodal point, are therefore not supported by the single-site approximation.
The observed difference indicates that accounting for the effect of paramagnons alone is insufficient to reproduce the two-gap structure in the electronic spectra.
To identify the missing ingredient, we note the key difference between the two methods.
The DF lattice DQMC scheme utilizes an exactly solved lattice reference system, which efficiently captures the ``bad fermion'' behavior that emerges already at rather high temperatures.
This behavior manifests itself as the simultaneous appearance of the pseudogap at the antinodal point in the electronic spectrum and the formation of LMMs in the system, composed of these ``bad'' electrons. 
These effects are entirely absent in the single-site impurity problem and cannot be fully accounted for by paramagnetic spin fluctuations within ladder-type diagrammatic approximations.
In the DF lattice DQMC scheme this effect exists already in the non-perturbative solution of the reference AFM Mott insulator system due to strong space-time correlations in the Monte-Carlo auxiliary spin-fields.

We note that the LMM in the single-band model for HTSC cuprates forms differently from that in multi-orbital systems such as iron pnictides~\cite{Kotliar2011}.
In the latter case, LMM formation is driven by strong Hund's exchange coupling~\cite{Georges_Hund}, making the static approximation for the LMM well justified.
In contrast, in the single-band Hubbard model, the LMM, associated with a single lattice site, does not exist in the static limit, and its emergence is not a true physical transition but rather a crossover~\cite{PhysRevB.105.155151}.
Nevertheless, LMM formation in the single-band case can be identified through spontaneous symmetry breaking associated with the formation of a short-range AFM ordering, which gives rise to a nonzero average value of the static AFM component of the Higgs field linked to the LMM.
This effect can be incorporated into the \mbox{D-TRILEX} framework by introducing an effective Higgs condensate of paramagnons, corresponding to a classical (static) AFM field $h_{\rm AFM}$, as detailed in Materials and Methods.
The Higgs field plays a crucial role in the formation of a pseudogap, as discussed in Refs.~\cite{Sachdev2018, Ferrero2018}.
Here, we further reveal its impact on the superconducting response.

In the right column of Figure~\ref{fig:PG}, we show the \mbox{D-TRILEX} results calculated in the presence of the Higgs condensate. 
The value of the field ${h_{\rm AFM}=0.485}$ is selected to ensure that the obtained results qualitatively align with those of the DF lattice DQMC scheme.
After accounting for the effect of the Higgs field, the momentum dependence of the Green's function closely matches the DF lattice DQMC results shown in the bottom panels of Figure~\ref{fig:HTSCview}. 
In particular, now the normal Green’s function $G$ exhibits a pseudogap at the antinodal point, which further confirms that the formation of LMMs and the emergent ``bad fermion'' behavior are intrinsically connected processes. 
The appearance of the pseudogap leads to a reduction in spectral weight in the anomalous Green's function $F$, causing the extrema of $F$ to shift from the antinodal point towards the nodal point.
The inclusion of the Higgs field also leads to another significant modification.
The momentum dependence of the anomalous self-energy $S$ changes from the cosine form of the applied superconducting field, resulting from the scattering on the spatial magnetic fluctuations (bottom left panel), to a pattern resembling the Green's function with a momentum shift of ${{\bf k} \to {\bf k} + Q}$, where ${Q = \{\pi, \pi\}}$.
This behavior arises from the AFM correlations between the LMMs, which generate an effective classical AFM mode that drives electronic scattering in the self-energy, as detailed in the Materials and Methods section.
We also observe that incorporating the effect of the Higgs field significantly enhances the superconducting response, leading to a substantial increase in the maximum values of the anomalous part of both the Green’s function $F$ and self-energy $S$.

This result might look surprising since the formation of a pseudogap typically suppresses superconductivity by reducing the spectral weight at the Fermi surface. 
To investigate the role of the pseudogap in the superconducting response, we perform an additional \mbox{D-TRILEX} calculation in which the contribution of the Higgs field is included only in the normal part of the self-energy $\Sigma$. 
As shown in the Supplemental Material, this results in a reduced superconducting response, consistent with expectations. 
Specifically, the maximum value of the anomalous self-energy $S$ decreases from ${\simeq0.03}$ (bottom left panel of Fig.~\ref{fig:PG}) to ${\simeq0.02}$, confirming that the formation of the pseudogap suppresses superconductivity. 
However, when the Higgs condensate is included in both the normal ($\Sigma$) and anomalous ($S$) self-energies (bottom right panel of Fig.~\ref{fig:PG}), the maximum value of the anomalous self-energy increases to ${\simeq 0.05}$.
This indicates that the AFM correlations of LMMs, which are responsible for the formation of the pseudogap, simultaneously enhance superconductivity.
By comparing the two cases with the pseudogap present, we conclude that the Higgs condensate contributes a bit more than 50\% to the superconducting response, in addition to the standard spin-fluctuation mechanism of electronic scattering on paramagnons. 
This estimation is consistent with the amount of mysterious contributions to the superconducting pairing found in C-DMFT calculations~\cite{Millis50}, establishing the LMMs formation and their AFM correlations as the missing strong-coupling mechanism of $d$-wave superconductivity. 
Note that bifacial effect of AFM correlations of LMMs on superconductivity is reminiscent of the role of plasmons, which suppress superconductivity by forming plasmaron shadow bands and reducing the spectral weight at the Fermi surface, while simultaneously enhancing it by providing additional glue for electron pairing~\cite{Veld2023}.

\section*{Materials and Methods}
\vspace{-7pt}
The $t$\,-\,$t'$ Hubbard model for single correlated cuprate band can be written as
\begin{align}
{\hat H_\alpha}=-\sum_{i,j,\sigma} t^{(\alpha)}_{ij} c_{{i}\sigma}^{\dagger} c_{{j}\sigma}^{\phantom{\dagger}}
+ \sum_{ i} U  n_{{i}\uparrow } n_{{i}\downarrow},
\label{Ham}
\end{align} 
where the operator $c_{{i}\sigma}^{(\dagger)}$ creates (destroys) an electron on a lattice site $i$ with the spin projection ${\sigma\in\{\uparrow,\downarrow\}}$. ${n_{{i}\sigma }= c_{{i}\sigma}^{\dagger} c_{{i}\sigma}^{\phantom{\dagger}}}$ is the density operator, $t^{\alpha}_{ij}$ is the hopping amplitude between the $i$ and $j$ sites, and $U$ is the on-site Coulomb repulsion.

We solve the introduced model using the so-called dual diagrammatic techniques pioneered by the Dual Fermion (DF) theory~\cite{RKL08}. 
This approach enables the treatment of essential correlation effects non-perturbatively by exactly solving a suitably chosen reference system. 
The remaining correlation effects, which extend beyond the reference problem, are then systematically included through a perturbative diagrammatic expansion.
The diagrammatic expansion is performed in the effective (dual) space for several key reasons.
First, it prevents double counting of correlations between the reference system and the remaining parts of the problem. 
Second, it transforms a non-perturbative at large $U$ expansion in terms of the original degrees of freedom into a perturbative one in the dual space. Consequently, the dual diagrammatic approach simultaneously combines weak- and strong-coupling expansions, making it exact in both these limits.
Furthermore, the dual diagrammatic expansion can be constructed based on arbitrary interacting reference systems, ranging from single~\cite{RKL08} or multiple~\cite{PhysRevB.97.115150, 10.21468/SciPostPhys.13.2.036, vandelli2024doping} impurity problems of dynamical mean-field theory~\cite{RevModPhys.68.13} to clusters of lattice sites~\cite{BRENER2020168310, Danilov2022, DFQMC}. 

{\bf Dual Fermion lattice DQMC scheme.}
In this work, we explore dual schemes that are based on two different reference systems. 
The first approach considers the ${16\times16}$ half-filled particle-hole-symmetric lattice as a reference. 
We introduce an $\alpha$ parameter to distinguish hopping amplitudes of the reference (${\alpha=0}$) and original (${\alpha=1}$) system. 
The hoppings can then be written in the following form:
\begin{align}
t^{(\alpha)}_{ij}=
\begin{cases}
t, & \text{for $(i, j) \in$ NN}\\
\alpha t', & \text{for $(i, j) \in$ NNN}\\
\alpha \mu, & \text{for $i=j$} \\
0, & \text{otherwise}
\end{cases}
\label{tij}
\end{align}
where $t$ is the nearest-neighbor (NN) and $t'$ is the next-nearest-neighbor (NNN) hopping amplitudes on a square lattice.
The chemical potential, $\mu$, is defined relative to that of the half-filled particle-hole symmetric case (${t'=0}$), where ${\mu_0=U/2}$.

\begin{figure}[t!]
\centering
\includegraphics[width=0.8\linewidth]{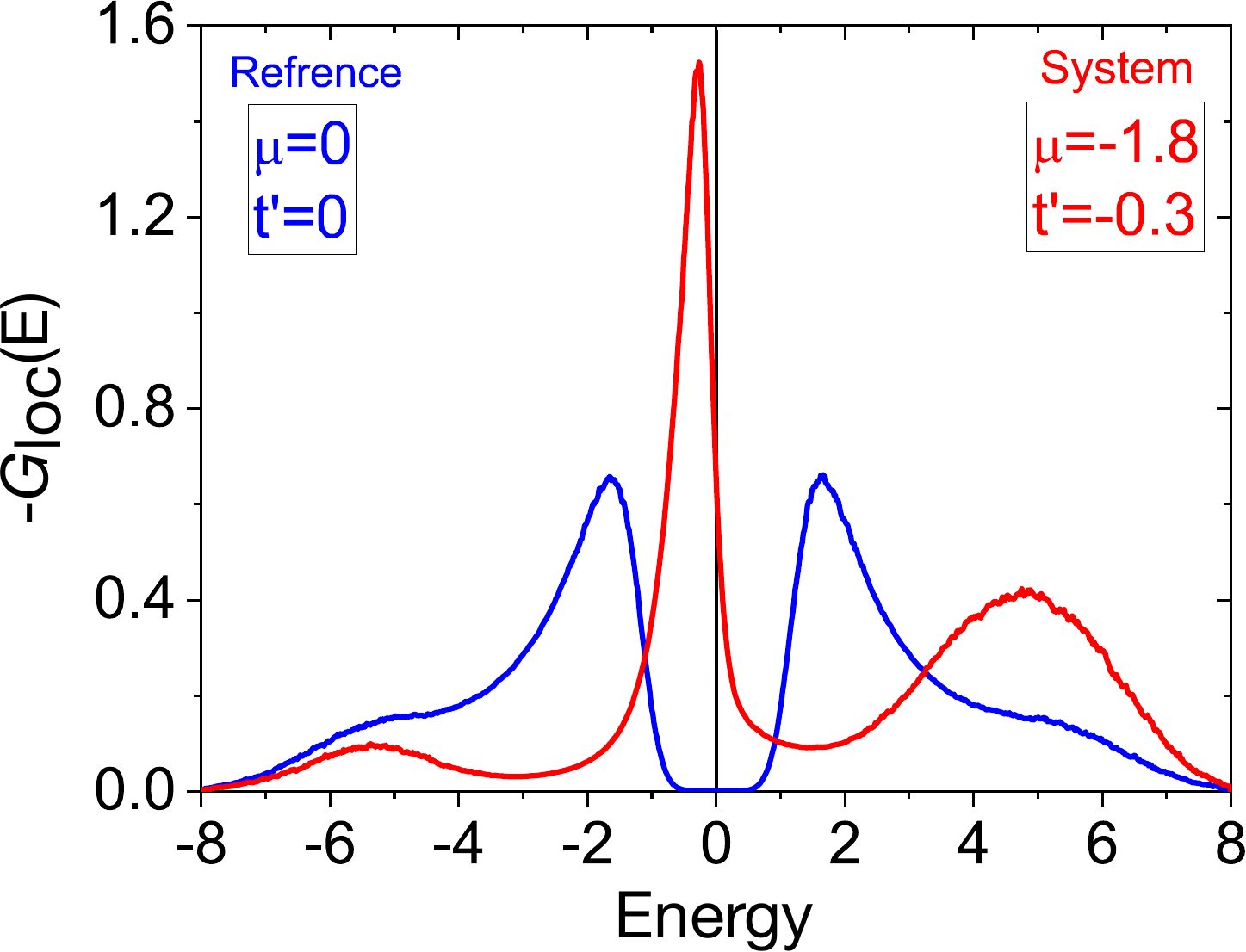}
\caption{Imaginary part of the local Green's function (proportional to the density of states) obtained for ${\beta=10}$ and ${U=5.6}$ for the half-field reference system (blue line) and the lattice problem with ${t'=-0.3}$ and ${\mu=-1.8}$ (red line), which corresponds to 13$\%$ hole doping.}
\label{fig:refQMC}
\end{figure}

The advantage of considering the specified reference problem is that it can be solved numerically exactly in the strongly correlated regime (${U\approx8t}$) using the Green's function auxiliary-field determinant quantum Monte Carlo (DQMC) method~\cite{DQMC_Scalettar}. 
This approach allows simulations on relatively large ${N \times N \times L}$ space-time lattices, with ${N\simeq20}$ sites and ${L\simeq 100}$ imaginary time slices, due to the absence of the fermionic sign problem~\cite{sign} in the particle-hole-symmetric case.
It is crucial for the present theoretical scheme that the reference system already incorporates a significant portion of the correlation effects in the system. If we look at the density of states proportional to local Green's function obtained by 
stohastic analitical continuation\cite{SOM2} for the half-filled Hubbard model for ${U=5.6}$ and ${t'=0}$ in Fig.~\ref{fig:refQMC}, one can see the existence of high-energy Hubbard bands as well as lower-energy antiferromagnetic Slater peaks at the Mott gap.
Such a ``four-peak'' structure is the characteristic feature of the half-filled Hubbard model for large interaction strength~\cite{Rost_QMC}. 
It is important to note that the Slater peaks cannot be captured by the paramagnetic single-site reference problem in the DMFT-like approximation~\cite{RKL08}.

To construct the diagrammatic expansion, we integrate out the reference problem and transform the fermionic degrees of freedom to the dual space, ${c^{(\dagger)} \to f^{(\dagger)}}$. This leads to the following action:
\begin{align}
\tilde{\cal S} = &-\Tr\sum_{1,2} \varphi^{*}_{1} \big[ {\cal \hat{\tilde{G}}} \big]^{-1}_{12} \varphi^{\phantom{*}}_{2} 
+ \tilde{\Phi}[f^{*},f],
\label{eq:dual_action}
\end{align}
written in terms of Grassmann variables $f^{(*)}$ for dual fermions.
The interaction part of the action ${\tilde{\Phi}[f^{*},f]}$ consists of the connected part of the exact four-point correlation function ${\langle c_{1,\sigma_1}^{\phantom{*}} c_{2,\sigma_2}^{*} c_{3,\sigma_3}^{\phantom{*}} c_{4,\sigma_4}^{*}\rangle}$ of the reference problem~\cite{RKL08, DFQMC}, where numbers represent the combined space-time $\{i,\tau\}$ indices. 
We perform calculations in the presence of an external superconducting $d$-wave field.
Therefore, it is convenient to work in the Nambu representation by introducing spinors
${\varphi^{*} = (f^{*}_{\uparrow}, f^{\phantom{*}}_{\downarrow})}$ and ${\varphi = (f^{\phantom{*}}_{\uparrow}, f^{*}_{\downarrow})^{T}}$.
In \eqref{eq:dual_action}, the bare dual Green's function 
\begin{align}
{\cal \hat{\tilde{G}}} =\big[\hat{\tilde{t}}^{-1} - \hat{g}\big]^{-1} = 
\begin{pmatrix}
\tilde{\cal G}^{\uparrow\uparrow} & \tilde{\cal G}^{\uparrow\downarrow}\\
\tilde{\cal G}^{\downarrow\uparrow} & -\tilde{\cal G}^{\downarrow\downarrow}
\end{pmatrix}
\label{eqn:Gd}
\end{align}
is a ${2 \times 2}$ pseudospin matrix in the Nambu space, and the trace is taken over this space.
In \eqref{eqn:Gd}, $g$ is the exact Green’s function of the reference system, and ${\tilde{t}} = t^{(1)} - t^{(0)}$ represents the difference between the hopping matrices including the chemical potentials as the diagonal term of the original and reference problems in Eq.(\ref{tij}), which serves as the perturbative parameter for the diagrammatic expansion.

The correlations effects beyond the reference problem are taken into account by the first-order contribution to the self-energy in the dual space~\cite{RKL08, DFQMC}: 
\begin{align}
\tilde{\Sigma }_{12}^{\uparrow \uparrow} &=-\sum_{ 3,4 }
\Bigl[
\langle{c_{1\uparrow}^{\phantom{*}}c_{2\uparrow}^{*}c_{3\uparrow}^{\phantom{*}} c_{4\uparrow}^{*}} \rangle \tilde{\cal G}^{\uparrow \uparrow}_{43}   +
\langle{c_{1\uparrow}^{\phantom{*}}c_{2\uparrow}^{*}c_{3\downarrow}^{\phantom{*}} c_{4\downarrow}^{*}} \rangle \tilde{\cal G}^{\downarrow \downarrow}_{43}   
\Bigr],
\notag \\
\tilde{\Sigma }_{12}^{\uparrow \downarrow} &=-\sum_{ 3,4 }
\langle{c_{1\uparrow}^{\phantom{*}}c_{2\downarrow}^{*}c_{3\downarrow}^{\phantom{*}} c_{4\uparrow}^{*}} \rangle \tilde{\cal G}^{\uparrow \downarrow}_{43},   
\label{df:Sig_spinor}
\end{align}
and similar for the remaining two spin components.

In order to develop a practical computational scheme for a large lattice system, we stochastically evaluate \eqref{df:Sig_spinor}.
Within the determinant DQMC scheme, which employs auxiliary Ising fields ${s}$, the Wick theorem can be applied to compute four-point correlation functions:
\begin{align}
 \langle{c_{1}^{\phantom{*}}c_{2}^{*}c_{3}^{\phantom{*}} c_{4}^{*}} \rangle _s =
 \langle{c_{1}^{\phantom{*}}c_{2}^{*}} \rangle _s \, 
 \langle{c_{3}^{\phantom{*}}c_{4}^{*}} \rangle_s - \langle{c_{1}^{\phantom{*}}c_{4}^{*}}\rangle_s \, 
 \langle{c_{3}^{\phantom{*}}c_{2}^{*}} \rangle _s
\label{gamma4qmc}
\end{align}
In order to subtract the disconnected part of the correlation function in \eqref{df:Sig_spinor}, we subtract the exact Green’s function of the reference system, ${g_{12} = -\langle c^{\phantom{*}}_{1} c_{2}^{\ast} \rangle}$, obtained from the previous DQMC run, as follows:
\begin{align}
\tilde{g}^s_{12}=g^{s}_{12}-g^{\phantom{s}}_{12}.
\label{gdfqmc}
\end{align}
We also utilize Fourier space for the efficient evaluation of \eqref{df:Sig_spinor}.
Within the DQMC framework, the lattice auxiliary Green’s function is not translationally invariant, meaning ${g^s_{12} = -\langle c_{1}^{\phantom{*}} c_{2}^{\ast} \rangle_s}$. 
To calculate $\tilde{g}^s_{kk'}$, we employ a double fast Fourier transform to momentum ${\bf k}$ and Matsubara frequency $\nu$ space, with ${k \in \{{\bf k}, \nu}\}$.
It is important to note that the dual Green's function, $\tilde{\cal G}^{\sigma \sigma'}_{k}$, is translationally invariant. Additionally, after performing the Monte Carlo summation over the auxiliary spins ${s}$, the self-energy also becomes translationally invariant. The final expression for the dual self-energies then reads:
\begin{align}
\tilde{\Sigma }_{k}^{\uparrow \uparrow} &=-\sum_{ s,k' } \Bigl[\left( 
\tilde{g}_{kk}^{\uparrow s} \tilde{g}_{k' k'}^{\uparrow  s }  
-\tilde{g}_{kk'}^{\uparrow s} \tilde{g}_{k'k}^{\uparrow s}    
\right) \tilde{\cal G}^{\uparrow \uparrow}_{k'} 
+\tilde{g}_{kk}^{\uparrow s} \tilde{g}_{-k', -k'}^{\downarrow s} \tilde{\cal G}^{\downarrow \downarrow}_{-k'} \Bigr],
\notag\\
\tilde{\Sigma }_{k}^{\uparrow \downarrow} &=-\sum_{ s,k' }\tilde{g}_{kk'}^{\uparrow s} \tilde{g}_{-k, -k'}^{\downarrow s}  \tilde{\cal G}^{\uparrow \downarrow}_{k'}. 
\label{1korderQMCspinor}
\end{align}
Note that only ${\cal \hat{\tilde{G}}}_k$ contains both normal and anomalous contributions. 
In contrast, $\tilde{g}^{\uparrow s}_{kk'}$ and $\tilde{g}^{\downarrow s}_{kk'}$, obtained for the paramagnetic reference system in the auxiliary spins, are diagonal in the spin space. 
There is a normalization factor associated with the number of QMC steps (noting that there is no sign problem for the reference system), along with two additional factors of ${\beta N^2}$ arising from the double Fourier transform and the summation over $k'$.

In superconducting calculations, we include a small external $d$-wave field: 
\begin{align}
\Delta_{\bf k} = 2 \, h_{dw} \, (\cos k_x - \cos k_y), 
\label{dw}
\end{align}
which explicitly enters the perturbation hopping matrix in the Nambu space 
\begin{align}
\hat{\tilde{t}}_{\bf k}(h_{dw}) =
\begin{pmatrix}
\tilde {t}_{\bf k} & \Delta_{\bf k} \\
\Delta_{\bf k} & -\tilde {t}_{-{\bf k}} 
\end{pmatrix}.
\label{tdw}
\end{align}
The final expression for the lattice Green's function of real fermions has the following matrix form~\cite{RKL08, DFQMC}:
\begin{align}
\hat{G}_{k}=\left[ \left( \hat{g}_k+{\hat{\tilde{\Sigma}}_k} \right)^{-1}-\hat{\tilde{t}}_{\bf k}\right]^{-1} =
\begin{pmatrix}
G_{k} & F_{k} \\
F^{*}_{k} & -G_{-k}
\end{pmatrix},
\label{DF_k}
\end{align}
where we introduced a shortened notations for the normal ${G=G^{\uparrow\uparrow}}$ and anomalous ${F=G^{\uparrow\downarrow}}$ Green's functions in the Nambu space.

An example of the calculated normal local Green's function for the realistic $t'$ and the $\mu$ shift is presented in Fig.~\ref{fig:refQMC} in comparison with the reference case. 
We can clearly see the formation of a narrow quasiparticle peak from the ``low Slater band'', while the Hubbard bands stay approximately on the same position due to the local nature of Mott-correlations. 
The intensity of the metallic upper Hubbard bands increases due to ``merging'' with the upper Slater band.

{\bf D-TRILEX scheme.}
The second computational scheme is based on the dual triply irreducible local expansion (\mbox{D-TRILEX}) method~\cite{PhysRevB.100.205115, PhysRevB.103.245123, 10.21468/SciPostPhys.13.2.036}. 
For this approach we use the paramagnetic single-site impurity problem of DMFT as a reference system with the same chemical potential $\mu$ as in the original lattice problem. 
In this case, it is convenient to work in the momentum and Matsubara frequency space representation.
The dual action in \mbox{D-TRILEX} is similar to \eqref{eq:dual_action}, but the perturbation hopping matrix that enters the bare dual Green's function~\ref{eqn:Gd} is given by the difference ${\tilde{t}^{\phantom{*}}_k = t^{(2)}_{\bf k} - \Delta^{\phantom{*}}_{\nu}}$ between the lattice dispersion $t^{(2)}_{\bf k}$, that includes both the NN and NNN hoppings, and the hybridization function $\Delta_{\nu}$ of DMFT.  
The interaction part of the action $\tilde{\Phi}$ exploits the partially-bosonized \mbox{D-TRILEX} form~\cite{PhysRevB.100.205115}, in which the interaction between fermions is mediated by charge density (${\varsigma = {\rm d}}$) and magnetic (${\varsigma = {\rm m} \in {x, y, z}}$) fluctuations, described by the bosonic fields $b^{\varsigma}$:
\begin{align}
\tilde{\Phi} = - \frac12\sum_{q,\varsigma}  b^{\varsigma}_{-q}
\left[\tilde{\cal W}^{\varsigma}_{q}\right]^{-1} b^{\varsigma}_{q}
+ \sum_{\substack{k,q,\varsigma,\\\sigma\sigma'}} \Lambda^{\varsigma}_{\nu\omega} f^{*}_{k\sigma}\sigma^{\varsigma}_{\sigma\sigma'} f^{\phantom{*}}_{k+q,\sigma'} b^{\varsigma}_{q}.
\label{eq:fbaction}
\end{align}
In this expression, $\tilde{\cal W}^{\varsigma}$ is the interaction renormalized by the polarization operator $\Pi^{\rm imp\,\varsigma}_{\omega}$ of the reference impurity problem in the corresponding channel $\varsigma$. 
$\Lambda^{\varsigma}_{\nu\omega}$ is the exact three-point vertex function of the single-site reference problem that describes the renormalized fermion-boson coupling. 
${\sigma^{\rm d} = \mathbb{I}}$, and $\sigma^{\rm m}$ are the corresponding Pauli matrices in the spin space.
We note, that a recent attempt to investigate the role of a non-local (momentum-dependent) three-point vertex function in the electron pairing through spin fluctuations was proposed in Ref.~\cite{2024arXiv241001705Y}.

The \mbox{D-TRILEX} approach self-consistently incorporates the effects of spatial collective electronic fluctuations through ladder-type diagrams for the self-energy and polarization operator in the dual space.
In the presence of the superconducting field, the self-energy develops an anomalous contribution, while the normal part of the self-energy remains the same as in the paramagnetic case~\cite{PhysRevB.100.205115, 10.21468/SciPostPhys.13.2.036}:
\begin{align}
\tilde{\Sigma}^{\uparrow\uparrow}_{k} 
= &-\sum_{q,\varsigma} 
\Lambda^{\varsigma}_{\nu\omega} \tilde{G}^{\uparrow\uparrow}_{k+q} \tilde{W}^{\varsigma}_{q} \Lambda^{\varsigma}_{\nu+\omega,-\omega}
+ 2\sum_{k'} 
\Lambda^{\rm c}_{\nu,0} \tilde{\cal W}^{\rm c}_{0} \Lambda^{\rm c}_{\nu',0} \, \tilde{G}^{\uparrow\uparrow}_{k'},
\notag\\
\tilde{\Sigma}^{\uparrow\downarrow}_{k} 
= &-\sum_{q,\varsigma} 
\xi^{\varsigma}\Lambda^{\varsigma}_{\nu\omega} \tilde{G}^{\uparrow\downarrow}_{k+q} \tilde{W}^{\varsigma}_{q} \Lambda^{\varsigma}_{-\nu,-\omega},
\label{eq:Sigma_DT}
\end{align}
where ${\xi^{\rm m/d} = \pm1}$.
Despite the Green's function has an anomalous part, the polarization operator remains diagonal in the channel space, but receives an additional term originating from the anomalous Green's function:
\begin{align}
\tilde{\Pi}^{\varsigma}_{q}  
= 2\sum_{k} \left(\Lambda^{\varsigma}_{\nu+\omega,-\omega} \tilde{G}^{\uparrow\uparrow}_{k} \tilde{G}^{\uparrow\uparrow}_{k+q} \Lambda^{\varsigma}_{\nu\omega}
+ \xi^{\varsigma} \Lambda^{\varsigma}_{-\nu,-\omega} \tilde{G}^{\downarrow\uparrow}_{k} \tilde{G}^{\uparrow\downarrow}_{k+q} \Lambda^{\varsigma}_{\nu\omega}\right).
\label{eq:dual_pol}
\end{align}
The dressed dual Green's function and the renormalized interaction are obtained via the corresponding Dyson equations:
\begin{align}
\big[\hat{\tilde{G}}_{k}\big]^{-1} &= \big[{\cal \hat{\tilde{G}}}_{k}\big]^{-1} - \hat{\tilde{\Sigma}}^{\phantom{*}}_{k},\\
\big[\tilde{W}^{\varsigma}_{q}\big]^{-1} &= \big[{\cal \tilde{W}}^{\varsigma}_{ q}\big]^{-1} - \tilde{\Pi}^{\varsigma}_{q},
\end{align}
and the lattice Green's function is obtained using \eqref{DF_k}. 
The normal ${\Sigma_{k} = \Sigma^{\uparrow\uparrow}_{k}}$ and anomalous ${S_{k} = \Sigma^{\uparrow\downarrow}_{k}}$ parts of the lattice self-energy can be obtained through the Dyson equation for the lattice Green's function.

To incorporate the effect of LMM formation in \mbox{D-TRILEX}, one has to assume spontaneous symmetry breaking, characterized by a non-zero condensate of the static (${\omega=0}$) AFM (${{\bf q}=Q=\{\pi,\pi\}}$) component of the spin-$z$ field in \eqref{eq:fbaction}:
\begin{align}
\langle b^{z}_{Q,0}\rangle = \pm h_{\rm AFM},
\end{align} 
which serves as the Higgs field~\cite{Sachdev2018, Ferrero2018, PhysRevB.105.155151} that introduces the AFM ordered LMMs in the system. 
Averaging the dual Green’s function, dressed by the Higgs field, over the positive and negative values of $h_{\rm AFM}$ effectively accounts for the singlet fluctuations and reproduces the paramagnetic case.
The averaging procedure results in an additional contribution to both the normal and anomalous parts of the self-energy:
\begin{align}
\hat{\tilde{\Sigma}}^{\rm AFM}_{k} = \Lambda^{z}_{\nu, 0}\hat{\tilde{G}}_{{\bf k}+Q,\nu}h^2_{\rm AFM}\Lambda^{z}_{\nu,0}.
\end{align}
In this expression $\hat{\tilde{\Sigma}}^{\rm AFM}$ and $\hat{\tilde{G}}$ are ${2\times2}$ matrices in the Nambu space.
We note that a similar contribution, but only to the normal self-energy, has been considered in Refs.~\cite{Sachdev2018, Ferrero2018}.
Recalling the \mbox{D-TRILEX} form for the self-energy~\ref{eq:Sigma_DT}, the derived LMM contribution to the dual self-energy can be accounted for by adding the following contribution to the $z$ component of the renormalized interaction:
\begin{align}
\tilde{W}^{z}_{q} \to \tilde{W}^{z}_{q} - h^2_{\rm AFM} \delta_{{\bf q},Q} \delta_{\omega,0}.
\end{align}
This additional term corresponds to an effective classical AFM mode and is similar to the one in Ref.~\cite{Irkhin1991}. 
Note that in our notations the renormalized interaction $\tilde{W}_{q}$ is negative.

In general, the non-zero value of the field $h_{\rm AFM}$ can be determined similarly to Ref.~\cite{PhysRevB.105.155151}, namely from the minimum of the free energy of the system at a temperature, where the free energy develops a double-well form as a function of the bosonic field ${b^{z}_{Q,0}}$.
In this work, we perform $\mbox{D-TRILEX}$ at a rather high temperature (${\beta = 10}$) and choose the value of the field ${h_{\rm AFM} = 0.485}$ to qualitatively reproduce the results obtained within the DF lattice DQMC scheme.\\

\begin{acknowledgements}
The authors thank Andy Millis, Viktor Harkov, Alexei Rubtsov, Igor Krivenko, Richard Scalettar, Emanuel Gull, Fedor \v{S}imkovic IV, Riccardo Rossi,  and Antoine Georges for valuable discussions. This work was supported by the Cluster of Excellence ``Advanced Imaging of Matter'' of the Deutsche Forschungsgemeinschaft (DFG) - EXC 2056 - Project No. ID390715994 and through the research unit QUAST, FOR 5249 - project No. ID449872909,  by European Research Council via Synergy Grant 854843 - FASTCORR.
\end{acknowledgements}

\bibliography{pnas}

\newpage

\onecolumngrid

\begin{center}
{\bf \large Supplemental Material\\[0.2cm]
Superconductivity of Bad Fermions: Origin of Two Gaps in HTSC Cuprates}
\end{center}


\twocolumngrid
\section{Matsubara Green's functions from DF-QMC}
\begin{figure}[t!]
\includegraphics[width=0.95\linewidth, trim={55pt 80pt 40pt 60pt}, clip]{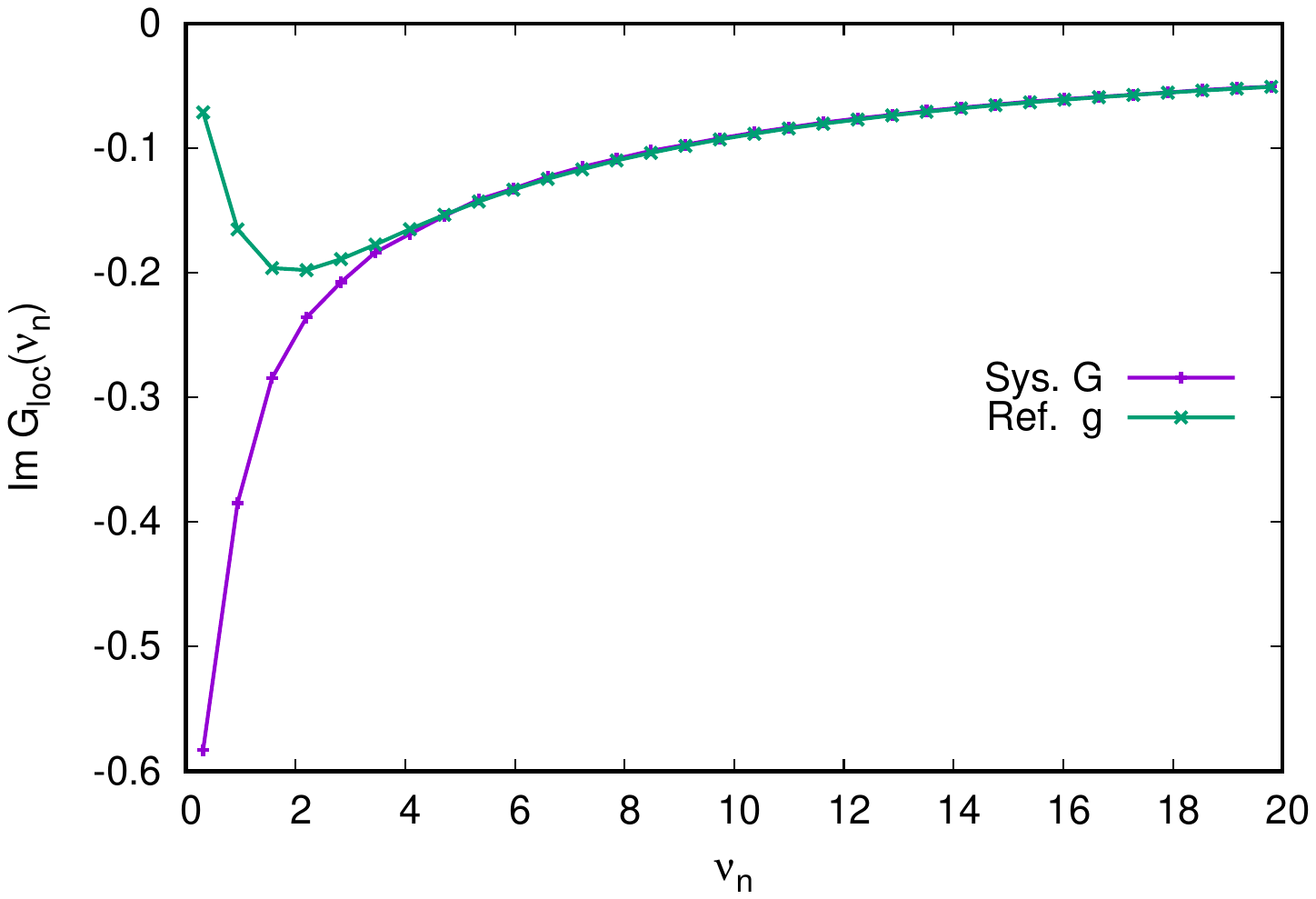}
\caption{The imaginary part of the normal local Green's function obtained for a ${16\times16}$ lattice at ${\beta=10}$, ${t=1}$, and ${U=5.6}$ using \mbox{DF-QMC}. 
The results are obtained for the half-filled reference $g$ (green) and lattice $G$ (magenta) problems, and plotted as a function of Matsubara frequency ${\nu_n=(2n+1)\pi/\beta}$.
The lattice problem is calculated for ${t'=-0.3}$ and ${\mu=-1.8}$ (${\delta=13.3\%}$ of hole doping). 
\label{fig:GlocB10}}
\end{figure}

\vspace{-7pt}
In this section, we present the Matsubara Green's functions as a ``raw'' output from the  DF-QMC scheme. 
The result is obtained for a  ${16\times16}$ periodic lattice.
In Fig.~\ref{fig:GlocB10}, we show the imaginary part of the local Green's function $g$ (green), as defined in Materials and Methods, obtained for the half-filled particle-hole-symmetric (${\mu=0}$, ${t'=0}$) reference problem using DQMC.
The reference Green's function is compared to the lattice Green's function $G$ (magenta) that is calculated within the DF-QMC scheme. 
One can see that in the high-frequency limit 
${\nu_n \ge 10}$ the dual perturbation is very small.
However, at low-frequency the lattice Green's function $G$ shows a metallic character, while the reference Green's function extrapolates to zero at ${\nu \to 0}$, which corresponds to the Mott-Slater gap. 
The stochastic analytical continuations of these Green's functions, shown in Fig.~(5) of the main text, confirms this behavior.

\begin{figure}[t!]
\includegraphics[width=0.95\linewidth, trim={55pt 80pt 40pt 60pt}, clip]{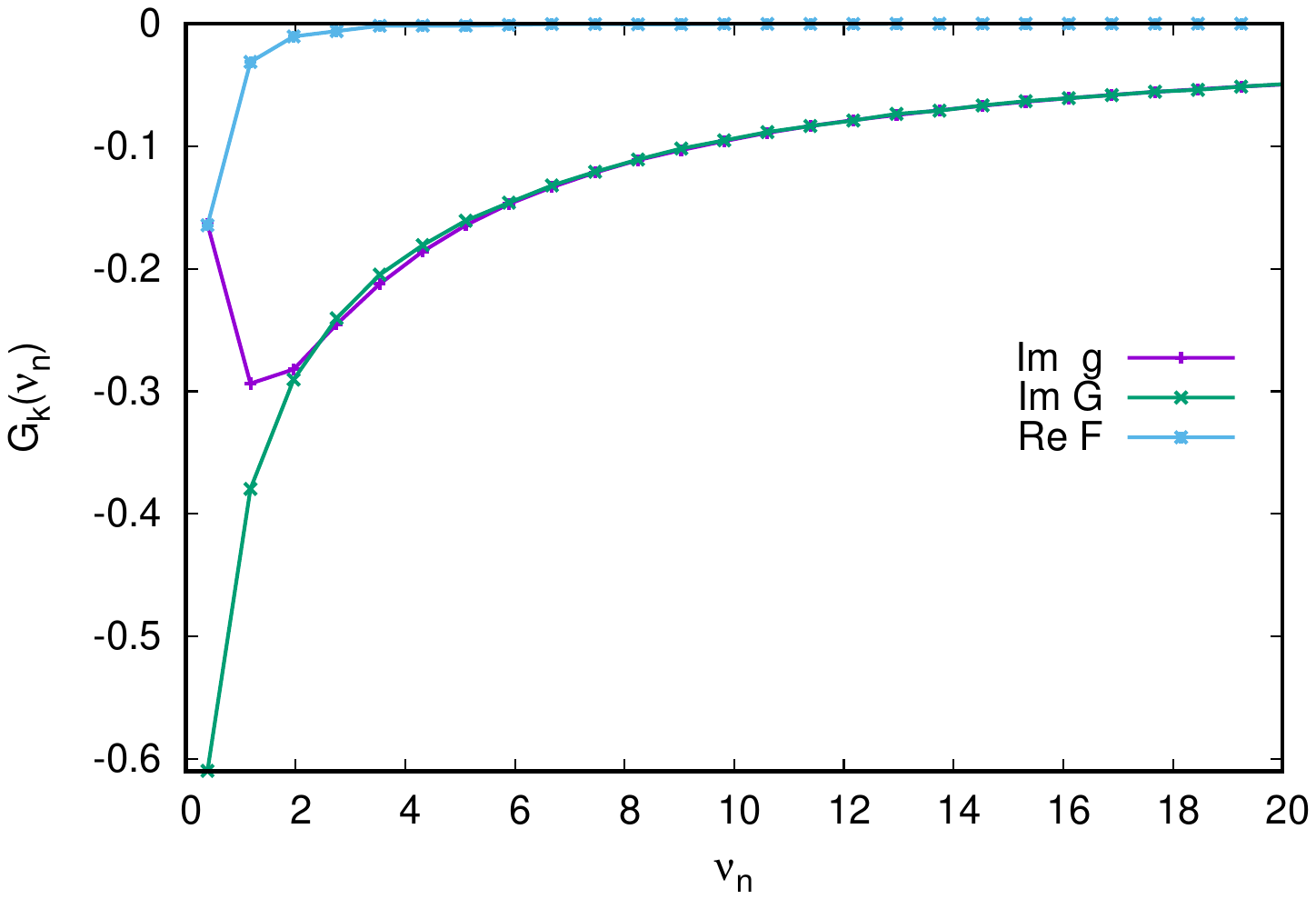}
\caption{The imaginary part of the normal Green's function of the reference $g$ (magenta) and lattice $G$ (green) problems, as well as the real part of the anomalous Green's function of the lattice problem $F$ (blue) obtained at ${\beta=8}$, ${t=1}$, and ${U=5.6}$ using \mbox{DF-QMC}.
The lattice problem is calculated for ${t'=-0.3}$ and ${\mu=-1.7}$ (${\delta=12\%}$ of hole doping) in the presence of an external superconducting field ${h_{dw}=0.05}$.
The results are obtained at ${\mathbf{k}=(\pi/4,3\pi/4)}$ momentum, which corresponds to maximum of $F({\bf k})$ on a ${16\times16}$ lattice, and plotted as a function of Matsubara frequency ${\nu_n=(2n+1)\pi/\beta}$.
\label{fig:GFB8}}
\end{figure}

To demonstrate the frequency dependence of the anomalous Green's function $F$, which is purely real in the Matsubara space, in Fig.~\ref{fig:GFB8} we show the DF-QMC result for $F({\bf k})$ (blue) calculated at ${\mathbf{k}=(\pi/4,3\pi/4)}$ momentum, which corresponds to maximum of $F({\bf k})$, as shown in the main text. 
The imaginary parts of the normal reference $g$ (magenta) and lattice $G$ (green) Green's functions are shown for comparison.

\begin{figure}[t!]
\includegraphics[width=1\linewidth, trim={55pt 60pt 40pt 60pt}, clip]{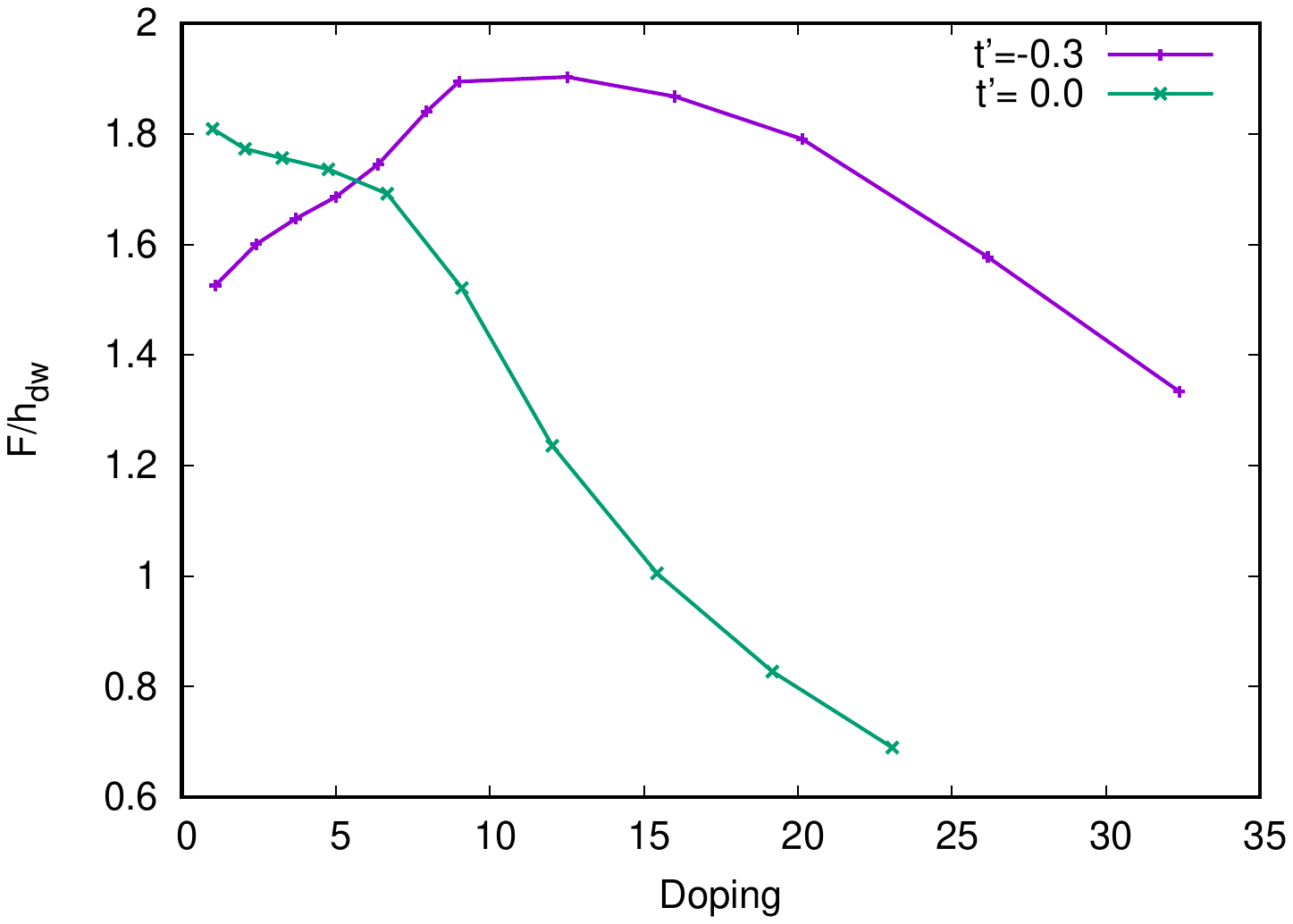}
\caption{The enhancement of the anomalous Green's function at the lowest Matsubara frequency ${\nu_0=\pi/\beta}$, with respect to the external $d$-wave field ${h_{dw}=0.05}$.
The ratio of $F/h_{dw}$ is calculated at the ${{\bf k}=X=(0,\pi)}$ point for the $8\times8$ lattice at ${\beta=10}$ using \mbox{DF-QMC} for difference NNN hoppings ${t'=-0.3}$ (magenta) and ${t'=0}$ (green).
\label{fig:Tc8x8}}
\end{figure}

We would also like to discuss the dependence of superconducting enhancement on hole doping. 
For these calculation we consider a ${8\times8}$ periodic lattice with a ``bare'' fermionic bath, which is introduced to reduce effects of the finite spectrum of small a system. 
In Fig.~\ref{fig:Tc8x8} we plot the $d$-wave enhancement of the anomalous Green's function compared to the external superconducting field ${h_{dw}}$ calculated at the X point, i.e. $F/h_{dw}$, as a function of hole doping for the two values of the NNN hopping 
(${t'=-0.3}$ and ${t'=0}$).
Remarkably, the dome-like superconducting behavior is found only for ${t' \ne 0}$. 
Moreover, the maximal superconducting enhancement around ${\delta=15\%}$ corresponds to ${\mu=-1.4}$, which is very close to the position of the van Hove singularity $4t'=-1.2$ in the non-interacting spectrum. 
The increase of the superconducting enhancement at ${\delta=15\%}$ for the optimal NNN hopping ${t'=-0.3}$ compared to ${t'=0}$ is about a factor of two.

\begin{figure}[t!]
\includegraphics[width=1\linewidth, trim={55pt 60pt 40pt 60pt}, clip]{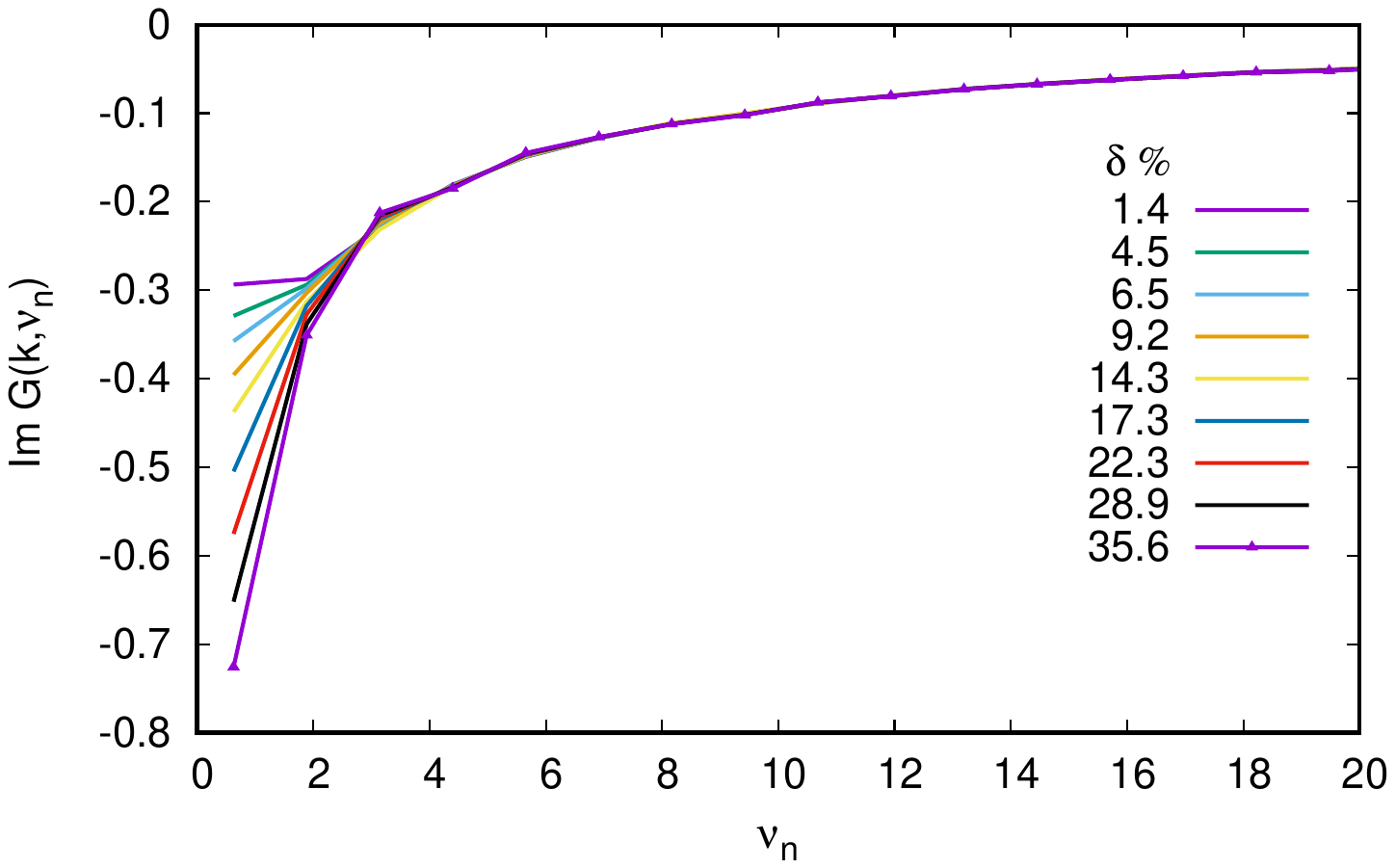}
\includegraphics[width=1\linewidth, trim={55pt 60pt 40pt 60pt}, clip]{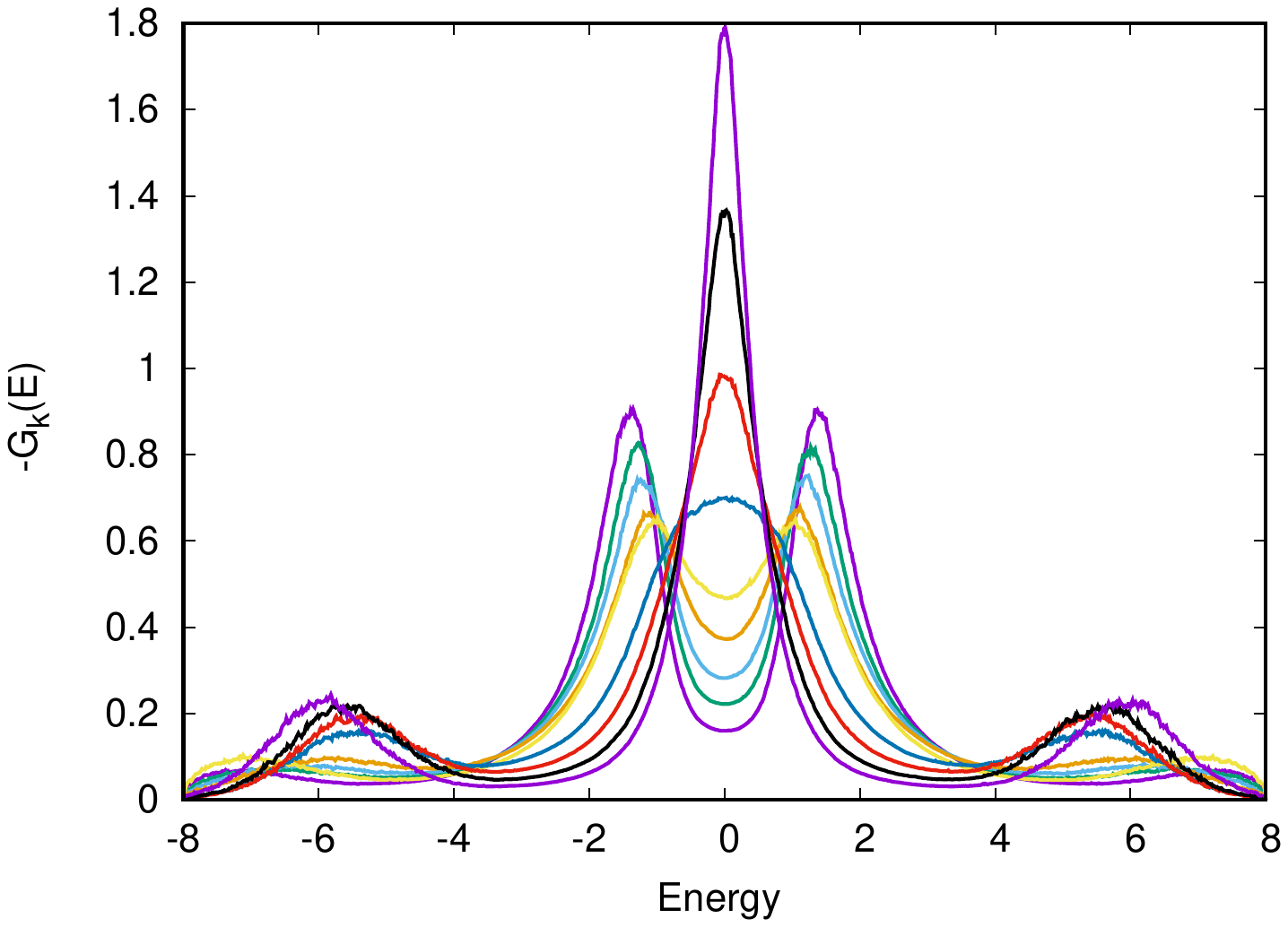}
\caption{The imaginary part of the normal Green's function on the Matsubara axis ${\nu_n=(2n+1)\pi/\beta}$ (top panel) obtained in \mbox{DF-QMC} for ${\beta=5}$, ${t=1}$, ${t'=-0.3}$ and ${U=5.6}$ and different chemical potentials $-\mu$=(0.8, 1.1, 1.2, 1.3, 1.45, 1.5, 1.6, 1.7, 1.8) which corresponds to the hole doping shows with different colors.
Corresponding analytical continuations using SOM-scheme on real axes -Im G$_k$(E) with the same color-scheme (bottom panel) for the ${{\bf k}=X=(0,\pi)}$.
\label{fig:GkE}}
\end{figure}

In order to investigate an interconnection of the pseudogap effects with coherence properties of "bad fermions" we calculate
the normal Matsubara Green's (Im G$_k$) within DF-QMC at the anti-nodal point for relative high temperature $\beta=5$ and different chemical potentials which
corresponds to underdoped system with $\delta=2-9 \%$, optimal doping $\delta=10-18 \%$, and overdoped system with $\delta=20-25\%$ (the top panel of the Fig.~\ref{fig:GkE}). It is useful to take the normalize trace of the Gor'kov Green's function
$(G_k-G_{-k})/2$ which just mean the particle-hole symmetry with the same (Im G$_k$) and (Re G$_k$=0). Using this Green's function,
we perform stochastic analytical continuation to the real energy (the bottom panel of the Fig.~\ref{fig:GkE}).
We can clearly seen the complicated transformation of the pseudogap spectral function for "bad-fermions" in underdoped system to more coherent fermions at optinally doped case with a DMFT-like three peak structure and finally to almost normal sharp correlated quasiparticles in the overdoped case.

\section{Effect of the Higgs field on the spectral function and superconductivity}
\vspace{-7pt}

\begin{figure}[t!]
\includegraphics[width=1\linewidth]{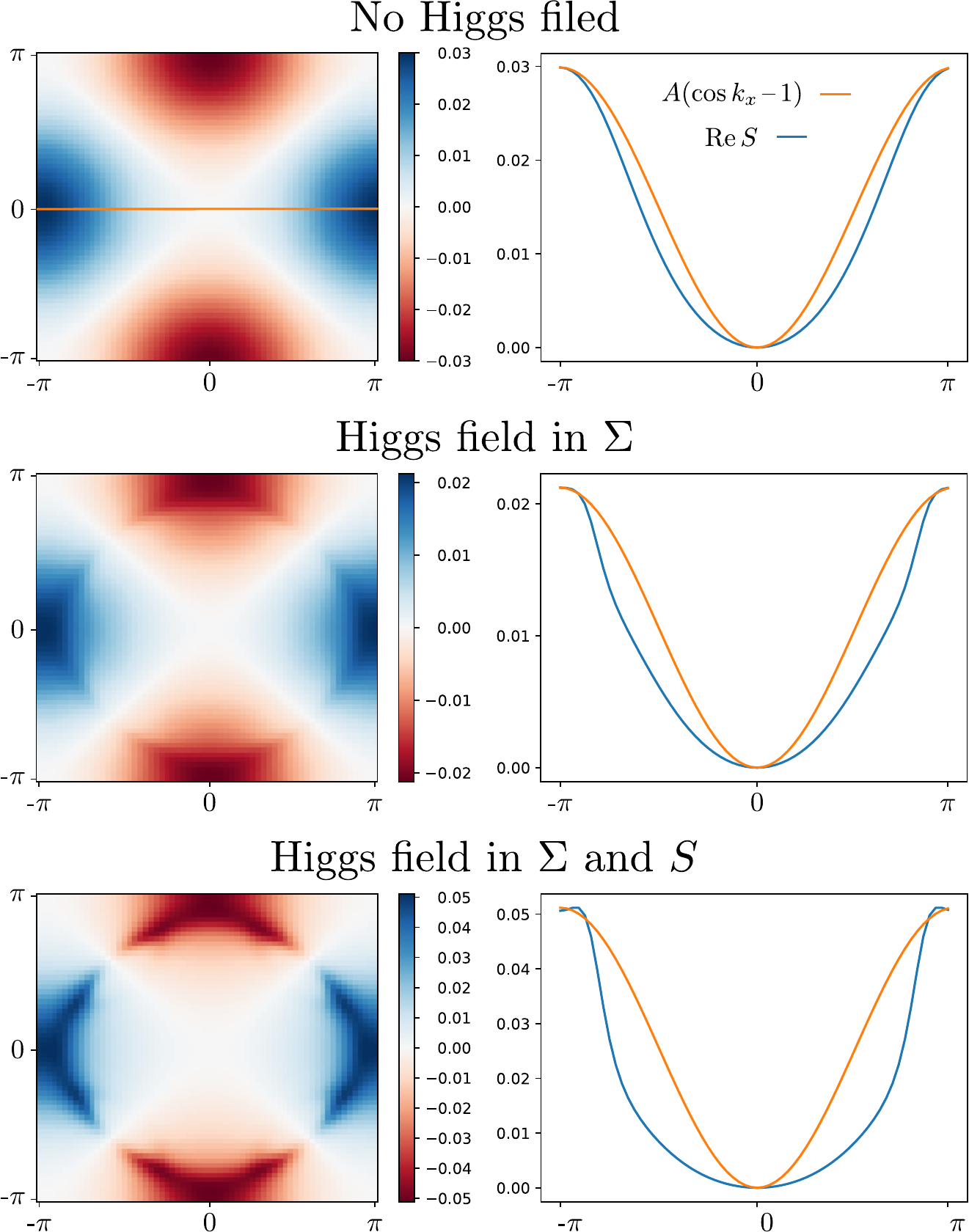}
\caption{The real part of the anomalous self-energy $S$ obtained using \mbox{D-TRILEX} at ${\beta=10}$ for ${\delta=16\%}$ of hole doping at the lowest Matsubara frequency ${\nu_0=\pi/\beta}$ as a function of momentum ${\bf k}$ in the full BZ (left column) and along the ${(k_x,0)}$ cut (right column) depicted in the top left panel by the horizontal orange line. 
The cut of the self-energy (blue curve) is compared to the cosine function typical to the $d$-wave field (orange curve). 
The results are calculated without the effective Higgs field (top panels), with the contribution of the Higgs field added only to the normal part $\Sigma$ of the self-energy (middle panels), and with the contribution of the Higgs field introduced in both normal $\Sigma$ and anomalous $S$ self-energies (bottom panels).
\label{fig:S_PG}}
\end{figure}

To investigate how AFM correlations of LMMs, introduced via the Higgs field, affect the spectral function and superconductivity, we compare the anomalous self-energies obtained within three different \mbox{D-TRILEX} calculations (see Materials and Methods). 
Fig.~\ref{fig:S_PG} shows momentum dependence of the real part of the anomalous self-energy $S$ calculated at the lowest Matsubara frequency ${\nu_0=\pi/\beta}$ at ${\beta=10}$ for ${\delta=16\%}$ hole doping.
The top row corresponds to the standard \mbox{D-TRILEX} calculation without introducing the Higgs field. 
In this case, the electronic spectral function does not reveal a pseudogap (see main text), and the form of the anomalous self-energy closely matches the ${(\cos k_x - \cos k_y)}$ function (orange line on the right panel) corresponding to the form of an applied external superconducting field with the $d_{x^2-y^2}$ symmetry.

To develop a pseudogap, that appears at the antinodal point of the electronic spectral function due to AFM correlations between the LMMs, we first include the contribution of the Higgs field only to the normal part $\Sigma$ of the self-energy and perform self-consistent \mbox{D-TRILEX} calculations.
The corresponding result for the anomalous self-energy is shown in the middle panels of Fig.~\ref{fig:S_PG}.
We find, that the formation of a pseudogap suppresses the superconduncting response, as the maximum value of ${\text{Re}\,S({\bf k})}$ reduces from ${\simeq0.03}$ to ${\simeq0.02}$ upon incorporating the effect of the Higgs field.
Additionally, we find that the form of the anomalous self-energy starts deviating from the cosine function and, in addition to the ${(\cos k_x - \cos k_y)}$ shape, displays enhanced intensities at momenta corresponding to the Fermi surface shifted by the AFM $Q$ vector, i.e. ${G({\bf k}+Q)}$.

We note that, as demonstrated in Materials and Methods, the AFM correlations of LMMs contribute not only to the normal $\Sigma$, but also to the anomalous $S$ part of the self-energy, through the scattering of electrons on an effective classical AFM mode that originates from the Higgs field.
Upon incorporating the effect of the Higgs field in both normal and anomalous self-energies in the self-consistent \mbox{D-TRILEX} calculation (bottom panels of Fig.~\ref{fig:S_PG}), we find that ${\text{Re}\,S({\bf k})}$ no longer follows the cosine form of the applied external $d$-wave field and even more resembles the Fermi surface, with a superimposed $d_{x^2-y^2}$ symmetry, shifted by the $Q$ vector.

More importantly, we observe that accounting for the effect of the Higgs field in both parts of the self-energy strongly enhances superconductivity. 
By comparing the maximum values of ${\text{Re}\,S({\bf k})}$ for the two cases, when the electronic spectral function reveals a pseudogap, shown on the middle and bottom panels of Fig.~\ref{fig:S_PG}, we find that AFM correlations between LMMs contribute for a bit more than 50\% to the superconducting response.

\end{document}